\documentclass[a4paper,11pt]{article}
\pdfoutput=1 

\usepackage[utf8]{inputenc}

\usepackage{jcappub} 

\usepackage[T1]{fontenc} 

\usepackage{epsf}
\usepackage{amsmath}
\usepackage{amssymb}
\usepackage{graphicx}  
\usepackage{epstopdf}
\usepackage{dcolumn}   
\usepackage{bm}        
\usepackage{color}
\usepackage{multirow}   
\usepackage{hyperref}
\usepackage[dvipsnames]{xcolor}
\usepackage{hyperref}
\hypersetup{colorlinks=true,urlcolor=blue,linkcolor=blue,citecolor=blue}

\usepackage{relsize}
\usepackage{soul}
\usepackage{caption}

\title{Checking the second law at cosmic scales}

\author[a]{Narayan Banerjee,} 
\author[b,1]{Purba Mukherjee,\note{Corresponding author}} 
\author[c]{Diego Pav\'{o}n}

\affiliation[a]{Department of Physical Sciences, Indian Institute of Science Education and Research Kolkata, Mohanpur, West Bengal - 741246, India.}
\affiliation[b]{Physics and Applied Mathematics Unit, Indian Statistical Institute, Kolkata - 700108, India.}
\affiliation[c]{Departamento de F\'{\i}sica, Facultad de Ciencias, Universidad Aut\'{o}noma de Barcelona, 08193 Bellaterra (Barcelona), Spain.}

\emailAdd{narayan@iiserkol.ac.in}
\emailAdd{purba16@gmail.com}
\emailAdd{diego.pavon@uab.es}

\abstract{Based on recent data about the history of the Hubble factor, it is argued that the second
	law of thermodynamics holds at the largest scales accessible to observation. This is consistent with
	previous studies of the same question.}


\arxivnumber{}

\begin{document}
\maketitle
\flushbottom

\section{Introduction \label{sec:1}}
\noindent The second law of thermodynamics is one of the pillars of our understanding the
physical world. According to this law, isolated macroscopic systems tend to a state of thermodynamic
equilibrium compatible with the constraints of the system itself. The equilibrium is characterized by a state
of maximum entropy, which means that, for isolated systems, the latter never decreases but eventually
attains a maximum at equilibrium. On the other hand, experience and common sense tell us that it will
never increase unbounded.

\noindent The common lore says that, nowadays, the entropy of the Universe (i.e., that
part of the Universe in causal contact with us) is strongly dominated by the entropy of
the cosmic horizon, $S_{h}$, nearly $10^{122}$ times the Boltzmann constant. The other main
sources of entropy (e.g. supermassive black holes, the cosmic background radiation, the cosmic
sea of neutrinos, etc.) contribute less by many orders of magnitude \cite{egan2010}.

\noindent As is well known, the second law is fulfilled at terrestrial scales and it is
routinely  applied at solar and galactic scales. Further, to the best of our knowledge,
all attempts to disprove it have failed. However, it remains to be seen if it is also satisfied
at the largest scales accessible to observation; i.e., at cosmic scales. Nevertheless, a
recent study suggests that this is indeed the case \cite{mnrs2019}. The target of this
research is to check whether the second law is satisfied at the aforementioned scales by a
different method that in Ref. \cite{mnrs2019}.  In the latter, three different and reasonable
parametrizations of the Hubble function were employed. By contrast we do not resort to any
parametrization; instead we use a well known expression of the current spatial curvature
parameter (Eq.  (\ref{eq:omegak0formula}), below) that holds at any time under consideration.

\noindent We resort to recent cosmological data on the history of the Hubble parameter;
namely,  cosmic chronometers (CC) measurements, baryon acoustic oscillations (BAO), and
luminosity distances from type Ia supernovae (SNIa) (32, 8 and 1624 data points, respectively).
These data are affected by the respective statistical and systematic
uncertainties whereby to obtain a sensible graph of the said history we apply
Gaussian processes \cite{rasmussen2005} to smooth the data, a widely used
technique in physical cosmology (see, e.g. \cite{seikel2012,shafieloo2012}).  Our only assumptions are
the validity of Einstein gravity and that at sufficiently large scales the Universe is
spatially homogeneous and isotropic, faithfully described by the Robertson-Walker metric.

\noindent Given that $S_{h} \propto {\cal A}$, where ${\cal A} = \frac{\pi}{H^{2} + k a^{-2}}$
is the area of the apparent horizon \cite{bak2000}, $a$ being the scale factor and  $H \equiv \dot{a}/a$
the Hubble function,  the second law implies that ${\cal A}$ cannot decrease with the expansion. As
it turns out we found that $d{\cal A}/da \geq 0$, or equivalently  $d{\cal A}/dz
\leq 0$ in the redshift range,  $ 0 < z \leq 2$, considered and thereby the second law is
fulfilled in the said range.

\noindent The manuscript is organized as follows. In section \ref{sec:2}, we
write the cosmic scales in terms of the deceleration parameter $q$ and curvature density
parameter $\Omega_{k}$. The methodology adopted is discussed in section \ref{sec:3}. Section
\ref{sec:4} introduces the observational data sets, followed by a brief description of
the Gaussian process algorithm, and the final reconstruction. We investigate the validity
of the second law at cosmic scales from the results in section \ref{sec:5}. Finally,
in section \ref{sec:6} we summarize our findings and make some concluding remarks.


\section{Theoretical Framework \label{sec:2}}

\noindent It is straightforward to see that the second law of
thermodynamics can alternatively be written as
\begin{equation}
	1+ q \geq \Omega_{k}\, ,
	\label{1plusq}
\end{equation}
where $q \equiv - \ddot{a}/(a H^{2})$
is the deceleration parameter and $\Omega_{k} \equiv - k/(a H)^{2}$
the spatial curvature parameter. Obviously this expression necessarily holds when the
universe is decelerating ($q >0$) or coasting ($q =0$),  since because of the Friedmann
equation $\mid \Omega_{k}\mid \leq 1$. However, for accelerating ($ q < 0$) the said expression
could be violated, at least in principle. Here we propose a novel experimental method to check
whether if this is the case.

\noindent From the above definition of spatial curvature it follows
\begin{equation}
	\Omega_{k}(z) = \Omega_{k0} \frac{(1+z)^{2}}{E^{2}(z)},
	\label{eq:defomegak}
\end{equation}
where $E(z) = H(z)/H_{0}$ is the normalized, dimensionless, Hubble factor.

\noindent Noting that $1+q = - \dot{H}/H^{2} = (1+z) \frac{H'}{H}$,
the second law can be recast as
\begin{equation}
	H'(z) \geq \Omega_{k0} \, H_0^{2} \, \frac{1+z}{H(z)},
	\label{eq:1+qOmegak}
\end{equation}
where a prime means differentiation with respect to redshift.

\noindent The latter sets a lower bound on the slope of the graph of
the Hubble function versus redshift. Here we wish to check the validity
of this inequality using updated observational data on the history of
the Hubble factor. These included Hubble parameter measurements from
CC, BAO, and luminosity distance measurements from SNIa.

\noindent The luminosity distance of any object (such as supernovae),
is given by
\begin{equation}
	d_{L}{}= \frac{c (1+z)}{H_{0} \sqrt{\mid \Omega_{k0}\mid}}\,  {\rm sinn} \left[\sqrt{\mid \Omega_{k0}\mid} \,
	\int_{0}^{z}{\frac{d\tilde{z}}{E(\tilde{z})}}\right] ,
	\label{eq:luminosity-distance}
\end{equation}
where  ${\rm sinn}[x]$ is $\sinh(x), x$ and $ \sin(x)$ for positive, zero and negative spatial curvature,
	respectively, and $c$ is the speed of light.
We define the transverse comoving distance to the source as,
\begin{equation}
	d_C(z) = \frac{d_{L}}{1+z},
	\label{eq:dz}
\end{equation}
and,
\begin{equation}
	D(z) = \frac{H_0}{c}d_C(z)
\end{equation}
is the normalized comoving distance.

\section{Methodology \label{sec:3}}

\noindent To investigate the validity of the second law of thermodynamics one needs to
obtain constraints on the parameter of spatial curvature. Equation  (\ref{eq:1+qOmegak})
amounts to the following condition
\begin{eqnarray}
	\zeta \equiv H'(z) - \Omega_{k0} \, H_0^2 \, \, \frac{1+z}{H(z)} \geq 0 , \label{eq:zeta}
\end{eqnarray}
for the non-violation of the second law of thermodynamics.

\noindent Now, given an estimate for the combination $\Omega_{k0}  H_0^2$, 
one can directly obtain $\zeta$ from Eq. \eqref{eq:zeta}.

\noindent Alternatively, one can
reconstruct the curvature density parameter by exploiting the relation
\begin{equation}
	\Omega_{k0} = \frac{E^{2}(z) \, D'^{2}(z) - 1}{D^{2}(z)} = \frac{H^{2}(z) \, d_C'^{2}(z) - c^2}{H_0^2 ~ d_C^{2}(z)},
	\label{eq:omegak0formula}
\end{equation}
which comes directly  from Eq. (\ref{eq:luminosity-distance}), see e.g. \cite{cai2015, yang2020}.

\noindent From equations (\ref{eq:1+qOmegak}), (\ref{eq:omegak0formula}) and the inequality, (\ref{eq:zeta}), we get
\begin{equation}
	\zeta \equiv H'(z) - \frac{(1 + z)}{H(z)} \, \frac{[H^{2}(z) \, d_C'^{2}(z) - c^2]}{d_C^{2}(z)} \geq 0.
	\label{eq:zeta_recon}
\end{equation}
This is independent of any specific cosmological model; it  solely rests on the Robertson-Walker metric.

\noindent The inequality (\ref{1plusq}) holds for decelerating and coasting universes. 
It is found to be satisfied at the present epoch for all types of spatial curvature 
($k= 0, +1, -1$) as well (see Mukherjee and Banerjee \cite{purba-prd2022}); however, it is not
guaranteed that it will hold at all intermediate redshifts. If experimentally it is
proved valid also between the commencement of the accelerated phase of cosmic
expansion (around a redshift of $0.78$, see e.g. \cite{daly2008})
and the present, it will strongly support the idea that the second law applies
to the observable universe and that the latter  behaves as a thermodynamic system.
As mentioned above, solid arguments in favor of this idea have been put forward in
Ref. \cite{mnrs2019} based on the history of the Hubble factor and
reasonable parametric expressions of the latter.

\noindent However, though we utilize $H(z)$ data as well, our method differs from theirs in that
we do not resort to any parametrization of the Hubble parameter. Instead we implement a
non-parametric reconstruction of the spatial curvature parameter (\ref{eq:omegak0formula}),
or directly utilize model-independent $\Omega_{k0}H_0^2$ constraints (\ref{eq:1+qOmegak})
from the Hubble and supernova datasets.

\section{Reconstruction \label{sec:4}}

\noindent Data-driven reconstruction is being widely used for model-independent predictions
in cosmology. Here, we resort to the machine-learning algorithm, Gaussian process (GP) 
regression \cite{seikel2012, shafieloo2012}, which can infer a function from a labeled set of training data.
Any function obtained from a GP is characterized by some mean and covariance function. The prior mean
function has been set to zero. So, the covariance function or `kernel', plays the central
role in encoding correlations between points where no data are available. Therefore, this
process can reproduce an ample range of behaviors and allows a Bayesian interpretation. A wide variety
of covariance functions exists in the literature. For this work, we consider six different kernels,
namely,

\noindent A. Squared Exponential (RBF)
\begin{equation}
	\kappa (z, \tilde{z}) = \sigma_f^2 \, \exp \left[ -\frac{(z - \tilde{z})^2}{2l^2}\right] \, ,
\end{equation}

\noindent B. Rational Quadratic (RQD)
\begin{equation}
	\kappa (z, \tilde{z}) = \sigma_f^2 \, \left[  1 + \frac{(z - \tilde{z})^2}{2 \alpha l^2} \right]^{-\alpha} \, ,
\end{equation}

\noindent C. Cauchy (CHY)
\begin{equation}
	\kappa (z, \tilde{z}) = \sigma_f^2 \, \left[  \frac{l}{(z - \tilde{z})^2 + l^2} \right] \, ,
\end{equation}

\noindent D. Mat\'ern with $\nu = \frac{5}{2}$ (M52), $\nu = \frac{7}{2}$ (M72), $\nu = \frac{9}{2}$ (M92)
\begin{equation}
	\kappa (z, \tilde{z}) = \sigma_f^2 \, \frac{{\left[\frac{\sqrt{2\nu}}{l}\, (z-\tilde{z})\right]}^{\nu}}
	{\Gamma(\nu) \, 2^{\nu -1}} \, K_\nu \left(\frac{\sqrt{2 \nu}}{l} (z-\tilde{z})\right)\, ,
\end{equation}
where $\nu$ denotes the order, $K_\nu(\cdot)$, and $\Gamma(\cdot)$ are the modified Bessel and Gamma functions.
The kernel hyperparameters, $\sigma_{f}$, $l$ and $\alpha$, typically control the strength of the fluctuations, 
the correlation length and the relative weighting of large-scale and small-scale variations at different length 
scales, between two separate redshifts.

\begin{figure*}[t!]
	\begin{minipage}{0.325\textwidth}
		\includegraphics[width=\textwidth]{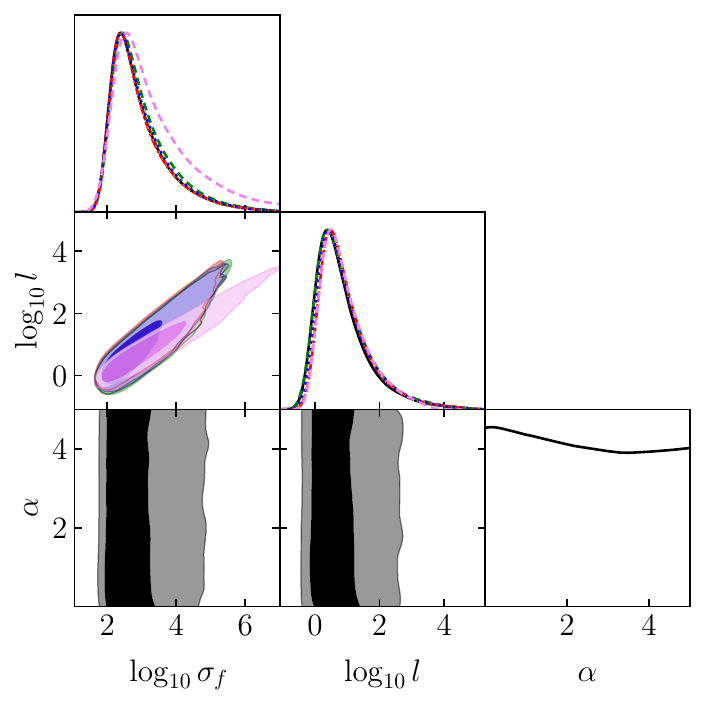}
		\caption*{(a) CC}
		\label{fig:contour_cc}
	\end{minipage}
	\begin{minipage}{0.325\textwidth}
		\includegraphics[width=\textwidth]{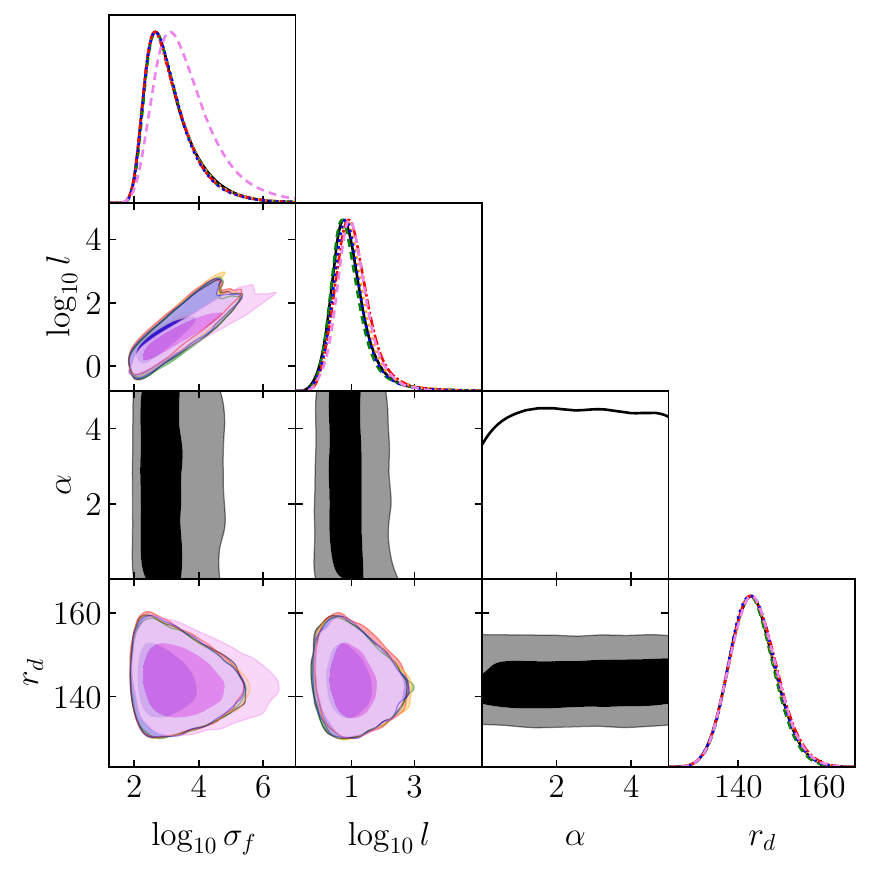}
		\caption*{(b) CC+BAO}
		\label{fig:contour_bao}
	\end{minipage}
	\begin{minipage}{0.325\textwidth}
		\includegraphics[width=\textwidth]{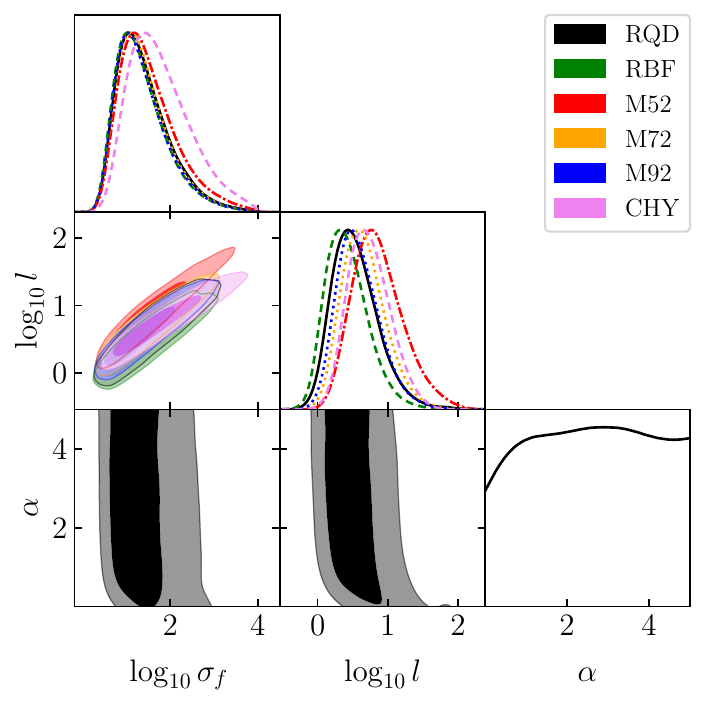}
		\caption*{(c) SN}
		\label{fig:contour_sn}
	\end{minipage}%
	\hfill 	
	\caption{Contour for GP hyperparameter space and the comoving sound horizon
		radius at drag epoch, $r_d$ [in units of Mpc], using different choices of the
		GP kernels.} 	\label{fig:hyper_plot}
\end{figure*}

\begin{table*}[t!]
	\begin{minipage}{0.5\linewidth}
		{\renewcommand{\arraystretch}{1.5} \setlength{\tabcolsep}{3 pt} \centering
			\begin{tabular}{l c c r}
				\hline \hline
				$\kappa(z, \tilde{z})$ &  $M$ & $\Omega_{k0} {h}^2$ &  $r_d$ [in units of Mpc] \\
				\hline
				RBF & $-19.407^{+0.100}_{-0.105}$ & $0.047^{+0.071}_{-0.070}$ & $143.203^{+5.547}_{-5.160}$  \\
				
				CHY & $-19.408^{+0.100}_{-0.105}$ & $0.054^{+0.070}_{-0.070}$ & $143.542^{+5.836}_{-5.389}$  \\
				
				RQD & $-19.405^{+0.099}_{-0.106}$ & $0.045^{+0.071}_{-0.070}$ & $143.348^{+5.722}_{-5.285}$  \\
				
				M52 & $-19.404^{+0.099}_{-0.104}$ & $0.041^{+0.069}_{-0.070}$ & $143.567^{+5.898}_{-5.359}$  \\
				
				M72 & $-19.408^{+0.100}_{-0.105}$ & $0.054^{+0.070}_{-0.070}$ & $143.360^{+5.741}_{-5.301}$  \\
				
				M92 & $-19.409^{+0.100}_{-0.106}$ & $0.054^{+0.071}_{-0.069}$ & $143.402^{+5.670}_{-5.238}$  \\
				
				\hline \hline
			\end{tabular}
		} \vskip -0.3cm
		\caption{{\small Constraints obtained on the parameters $M$, $\Omega_{k0} h^2$ and $r_d$ from different observational data.}}	
		\label{tab:constraints}
	\end{minipage} \hfill
	\begin{minipage}{0.4\linewidth}
		\includegraphics[width=0.925\textwidth]{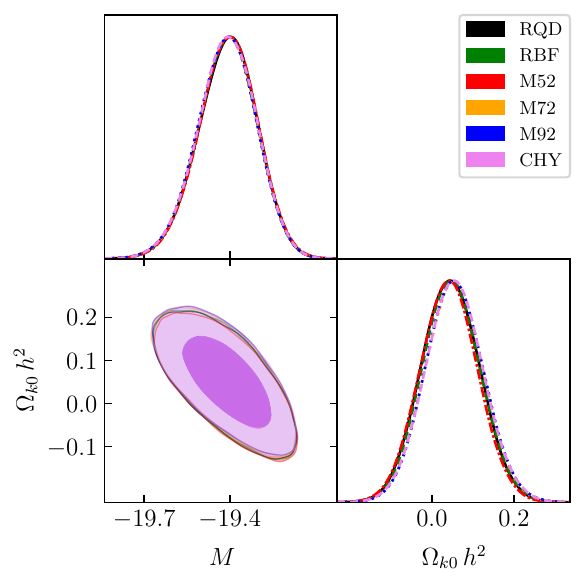} \\
		\captionof{figure}{Contour plots for $M$ and $\Omega_{k0} h^2$ using different GP kernels.} \label{fig:M_contour}
	\end{minipage}
\end{table*}

\begin{table*}[t!]
	\begin{center}
		\resizebox{\textwidth}{!}{\renewcommand{\arraystretch}{1.2} \setlength{\tabcolsep}{18 pt} \centering
			\begin{tabular}{l c c c c c c }
				\hline \hline
				Datasets & RBF & CHY & RQD & M52 & M72 & M92\\
				\hline
				CC &  0.419 & 0.398 & 0.409 & \textbf{0.384} & 0.401 & 0.408 \\
				
				CC+BAO &  0.403 & 0.401 & 0.402 & \textbf{0.397} & 0.400 & 0.401 \\
				
				SN &  0.311 & 0.310 & 0.310 & \textbf{0.305} & 0.308 & 0.309 \\
				
				\hline \hline
			\end{tabular}
		}
	\end{center}
	\caption{{\small Reduced $\chi^2$ obtained with different data sets for the following kernels. The minimum values are highlighted in bold}}
	\label{tab:chi2r}
\end{table*}

\begin{figure*}[t!]
	\begin{minipage}{\textwidth}
		\begin{minipage}{0.45\textwidth}
			\includegraphics[width=\textwidth, height=0.23\textheight]{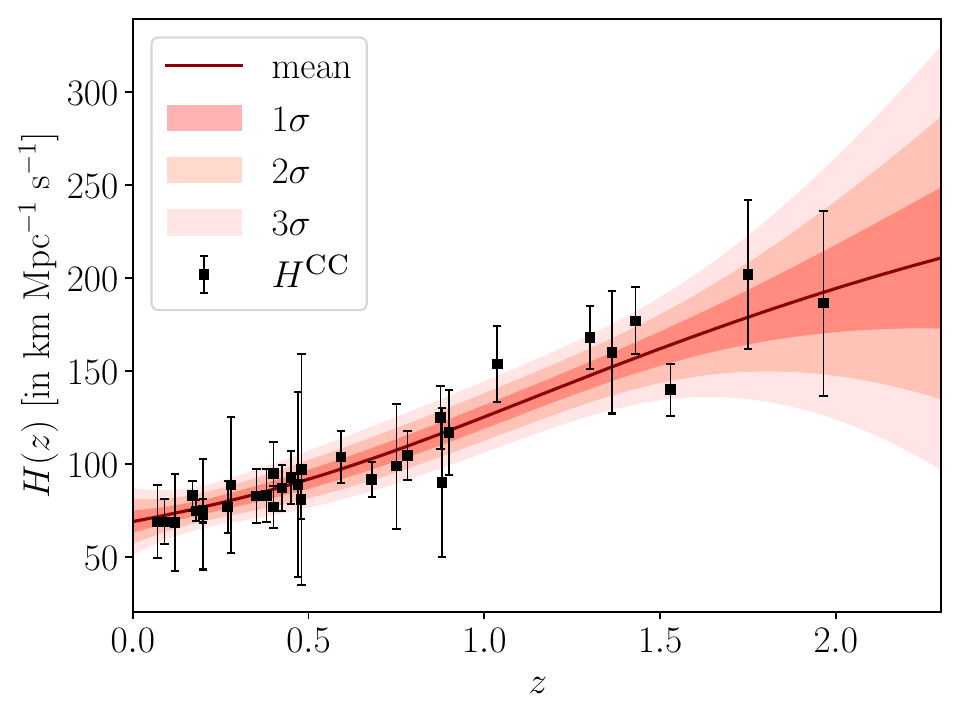}
		\caption*{(a) Reconstructed $H(z)$}
		\end{minipage}
		\begin{minipage}{0.45\textwidth}
			\includegraphics[width=\textwidth, height=0.23\textheight]{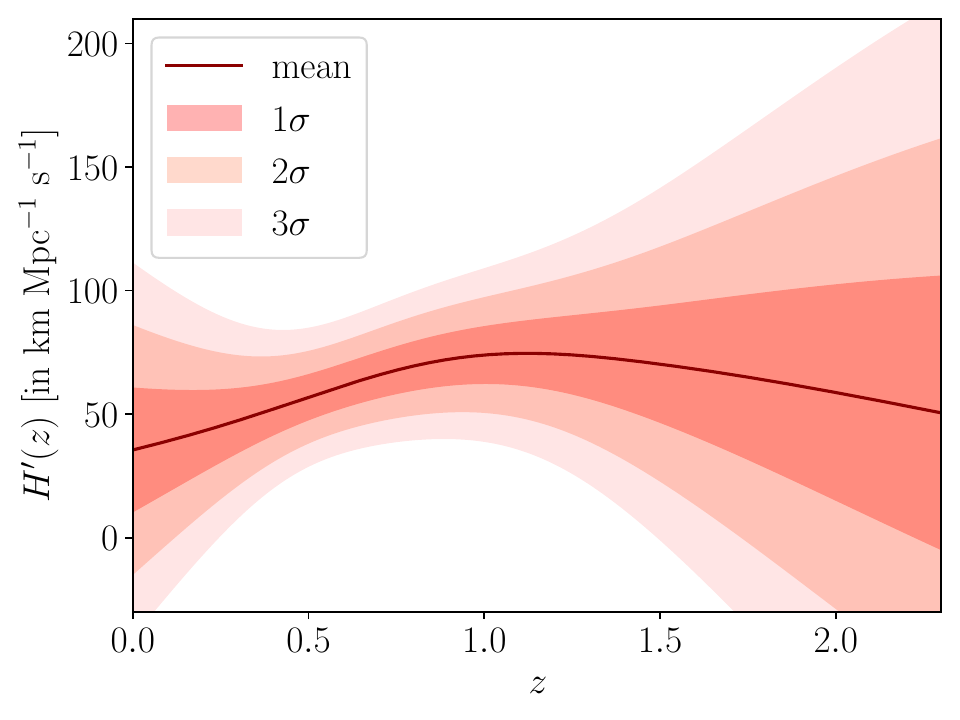}
			\caption*{(b) Reconstructed $H'(z)$}
		\end{minipage} ~~~~
	\end{minipage}
	\vskip 0.2cm
	\begin{minipage}{\linewidth}
		\begin{minipage}{0.325\textwidth}
			\includegraphics[width=\textwidth]{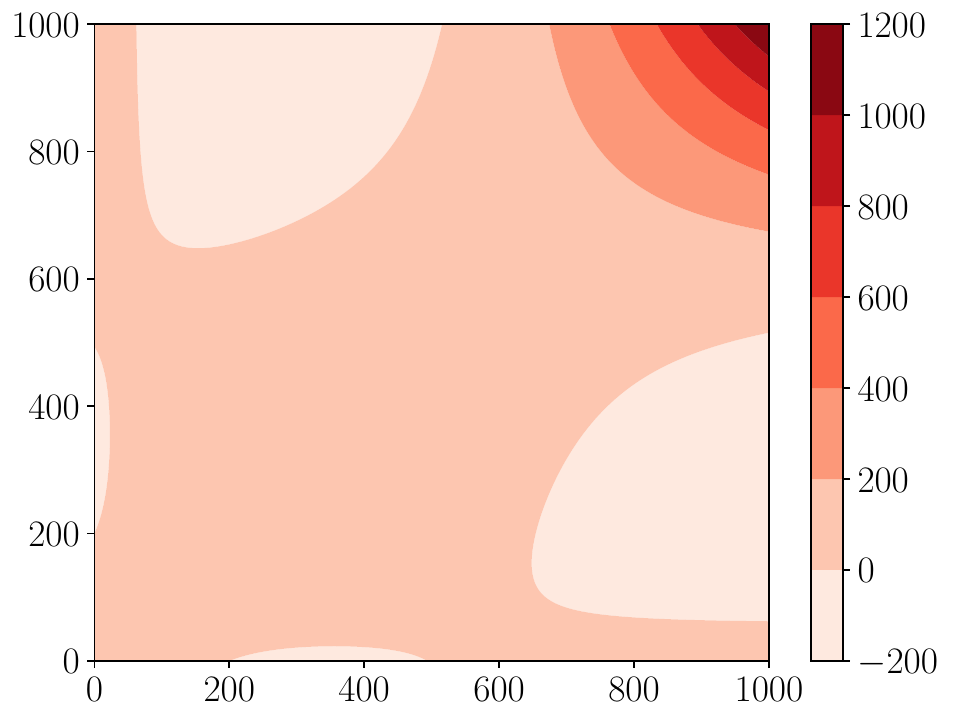}
			\caption*{(c) cov[$H(z_i)$, $H(z_j)$]}
		\end{minipage}%
		\begin{minipage}{0.325\textwidth}
			\includegraphics[width=\textwidth]{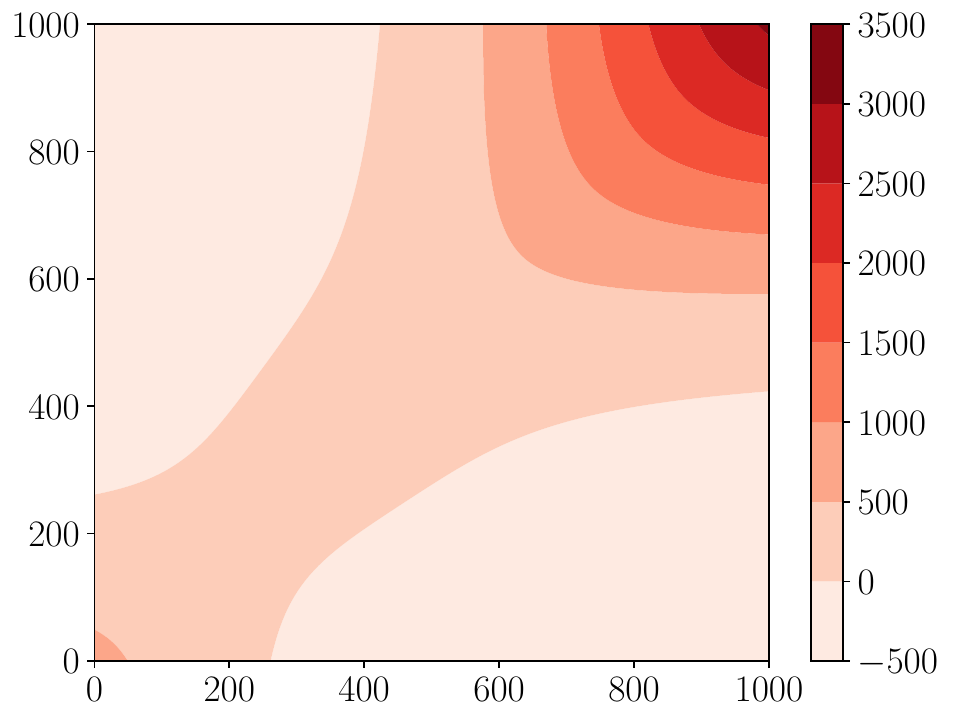}
			\caption*{(d) cov[$H'(z_i)$, $H'(z_j)$]}
			\end{minipage}%
		\begin{minipage}{0.325\textwidth}
			\includegraphics[width=\textwidth]{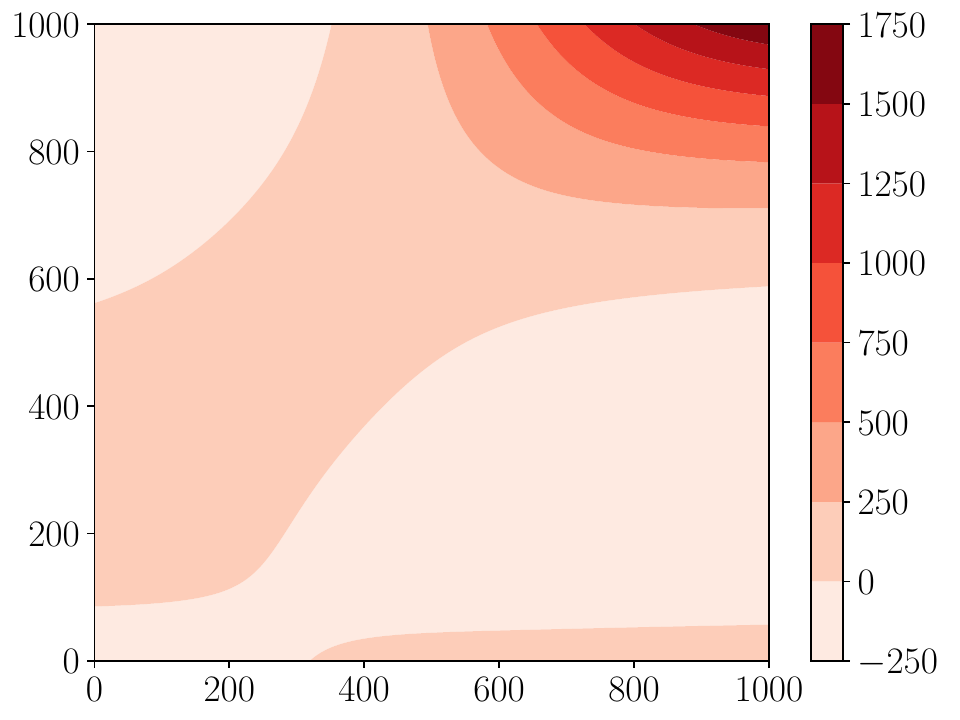}
			\caption*{(e) cov[$H(z_i)$, $H'(z_j)$]}
		\end{minipage}%
	\end{minipage}%
	\hfill 	
	\caption{Reconstructed functions (a) $H(z)$, (b) $H'(z)$ [in units of km Mpc$^{-1}$ s$^{-1}$] along with their $1\sigma$, $2\sigma$ and $3\sigma$ uncertainties using CC Hubble data set. Plots (c), (d) and (e) show the respective covariances between the predicted functions at different test redshifts.}
	\label{fig:H_rec}
\end{figure*}

\begin{figure*}[t!]
	\begin{minipage}{\textwidth}
		\begin{minipage}{0.45\textwidth}
			\includegraphics[width=\textwidth, height=0.23\textheight]{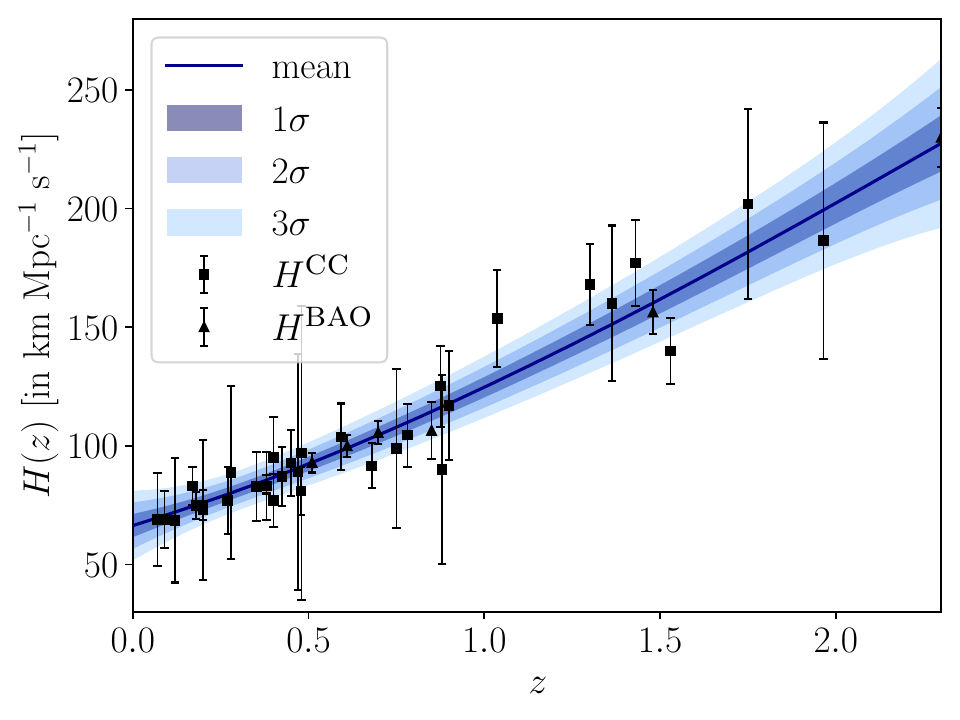}
			\caption*{(a) Reconstructed $H(z)$}
		\end{minipage}
		\begin{minipage}{0.45\textwidth}
			\includegraphics[width=\textwidth, height=0.23\textheight]{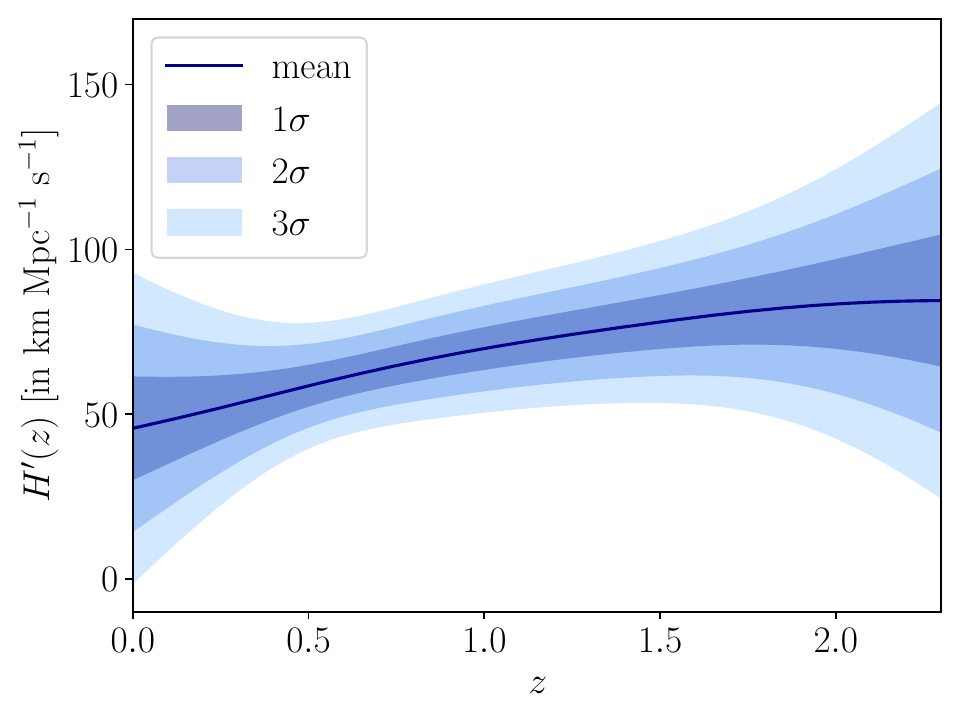}
			\caption*{(b) Reconstructed $H'(z)$}
		\end{minipage} ~~~~
	\end{minipage}
	\vskip 0.2cm
	\begin{minipage}{\linewidth}
		\begin{minipage}{0.325\textwidth}
			\includegraphics[width=\textwidth]{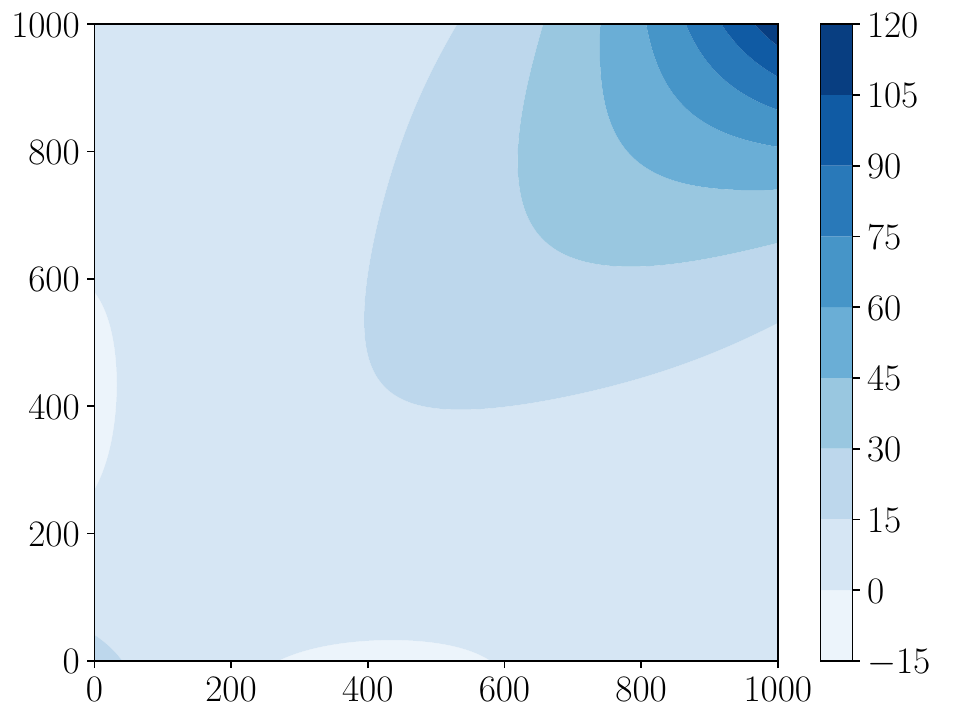}
			\caption*{(c) cov[$H(z_i)$, $H(z_j)$]}
		\end{minipage}%
		\begin{minipage}{0.325\textwidth}
			\includegraphics[width=\textwidth]{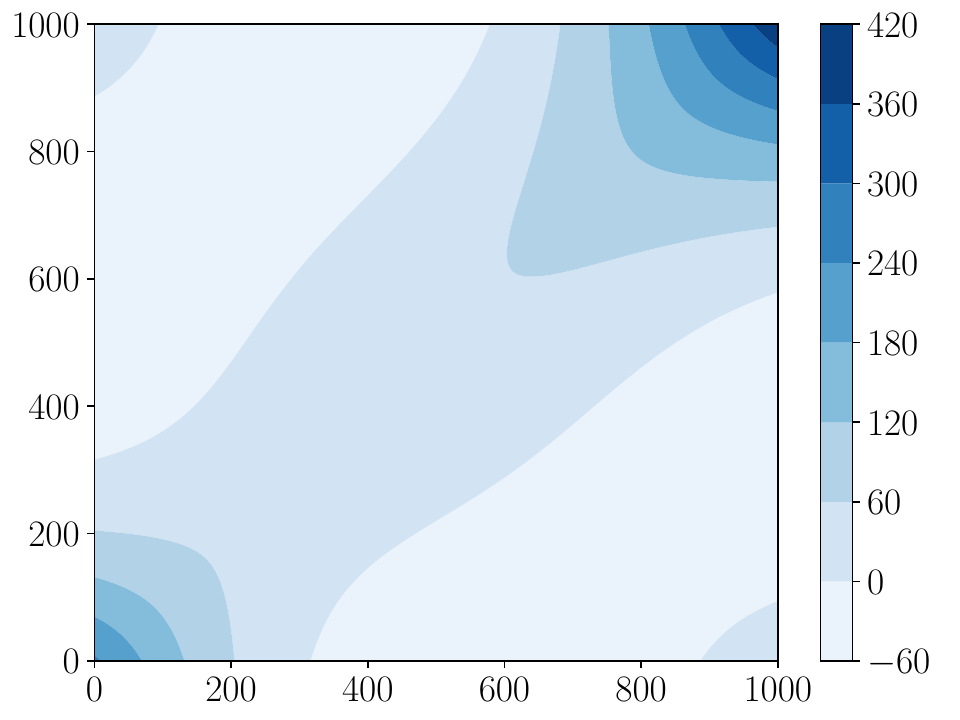}
			\caption*{(d) cov[$H'(z_i)$, $H'(z_j)$]}
		\end{minipage}%
		\begin{minipage}{0.325\textwidth}
			\includegraphics[width=\textwidth]{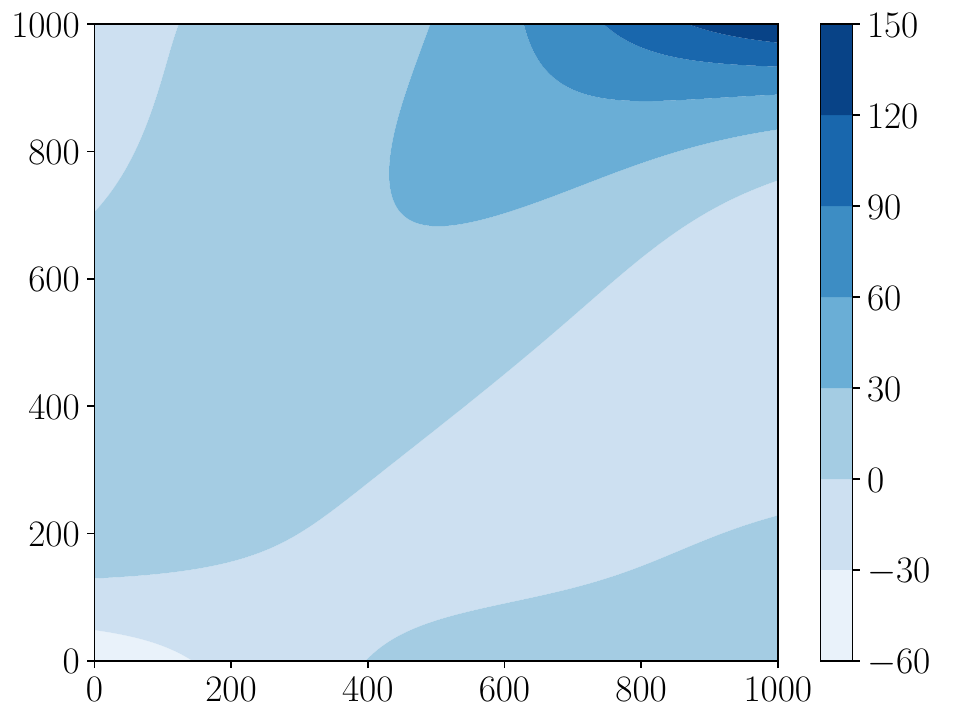}
			\caption*{(e) cov[$H(z_i)$, $H'(z_j)$]}
		\end{minipage}%
	\end{minipage}%
	\hfill 	
	\caption{Reconstructed functions (a) $H(z)$, (b) $H'(z)$ [in units of km Mpc$^{-1}$ s$^{-1}$] along with their $1\sigma$, $2\sigma$ and $3\sigma$
		confidence levels using CC+BAO Hubble data set. Plots (c), (d) and (e) show the respective covariances between the predicted functions at different test redshifts.}
	\label{fig:H2_rec}
\end{figure*}

\begin{figure*}[t!]
	\begin{minipage}{\textwidth}
		\begin{minipage}{0.45\textwidth}
			\includegraphics[width=\textwidth, height=0.23\textheight]{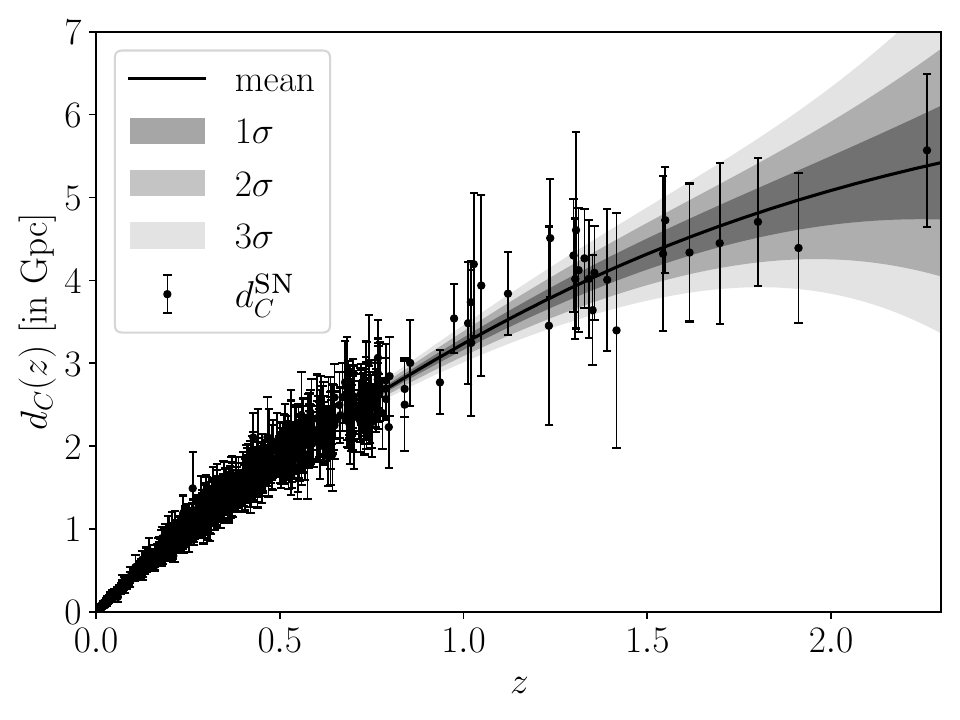}
			\caption*{(a) Reconstructed $d_C(z)$}
		\end{minipage}
		\begin{minipage}{0.45\textwidth}
			\includegraphics[width=\textwidth, height=0.225\textheight]{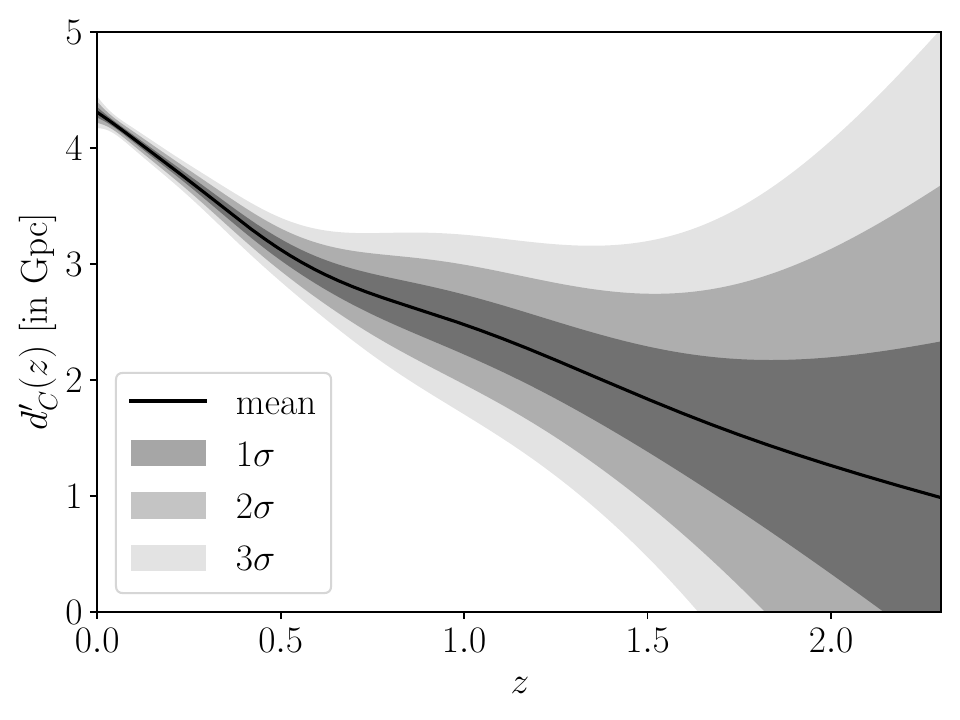}
			\caption*{(b) Reconstructed $d_C'(z)$}
		\end{minipage} ~~~~
	\end{minipage}
	\vskip 0.2cm
	\begin{minipage}{\linewidth}
		\begin{minipage}{0.325\textwidth}
			\includegraphics[width=\textwidth]{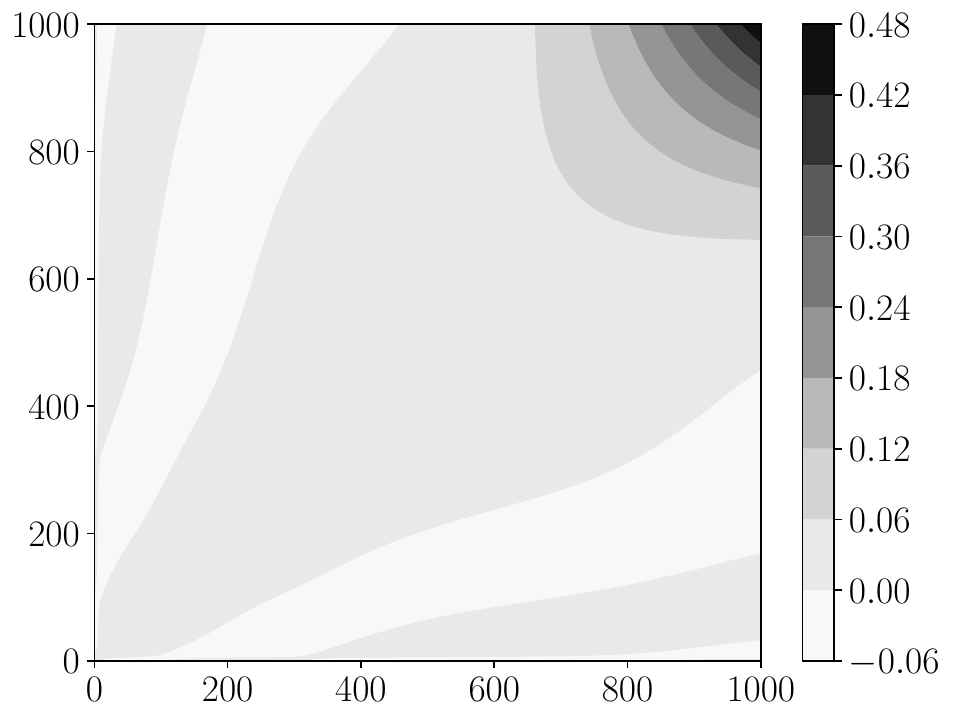}
			\caption*{(c) cov[$d_C(z_i)$, $d_C(z_j)$]}
		\end{minipage}%
		\begin{minipage}{0.325\textwidth}
			\includegraphics[width=\textwidth]{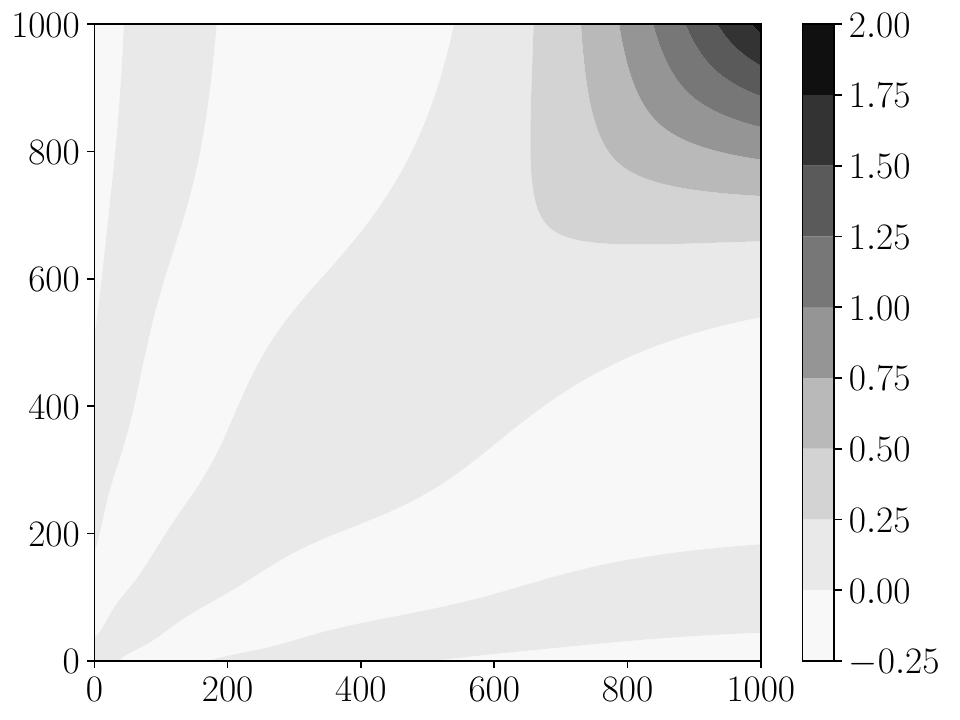}
			\caption*{(d) cov[$d_C'(z_i)$, $d_C'(z_j)$]}
		\end{minipage}%
		\begin{minipage}{0.325\textwidth}
			\includegraphics[width=\textwidth]{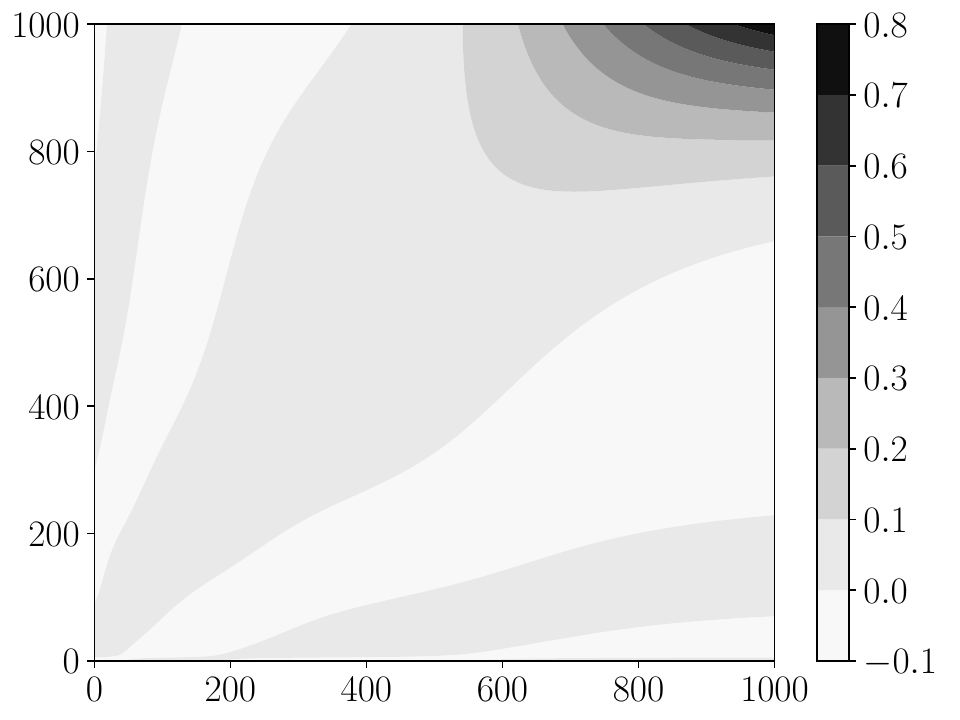}
			\caption*{(e) cov[$d_C(z_i)$, $d_C'(z_j)$]}
		\end{minipage}%
	\end{minipage}%
	\hfill 	
	\caption{Reconstructed functions (a) $d_C(z)$, (b) $d_C'(z)$ [in units of Gpc] along with their $1\sigma$,
		$2\sigma$ and $3\sigma$ confidence intervals using Pantheon+ SN Ia data set. Plots (c), (d) and (e) show the respective covariances
		between the predicted functions at different test redshifts.}
	\label{fig:dc_rec}
\end{figure*}

\noindent To begin with, we undertake a Gaussian process reconstruction of the Hubble
parameter $H(z)$ and its derivative $H'(z)$ from the recent 32 CC and combined
32 CC + 8 BAO data. The CC Hubble data depend on the differential ages
of galaxies and do not assume any particular cosmological model \cite{cc1,cc2,cc3,cc4,cc5,cc6,cc7,cc8}. 
We also take into account the systematic uncertainties reported in Moresco et al. \cite{cc_sys}. Again,
the BAO measurements \cite{bao1,bao2,bao3,bao4,bao5,bao6,bao7,bao8} make use of a fiducial radius of the
comoving sound horizon at drag epoch, $r_d^{\text{fid}}$, which depend on the underlying
cosmology. We adopt a full marginalization over the GP hyperparameter space setting $r_d$
for the BAO data as an additional free parameter to obtain a model-independent Hubble
data set from CC and the calibrated BAO.

\noindent We adopt a Bayesian Markov chain Monte Carlo (MCMC) analysis using the emcee \cite{emcee},
assuming flat priors on the kernel hyperparameters. The two dimensional
confidence contours showing the uncertainties along with the one dimensional marginalized
posterior probability distributions are shown in Fig. \ref{fig:hyper_plot} for CC (left
panel) and combined CC+BAO (middle panel) $H(z)$ reconstructions with GetDist \cite{getdist}.

\noindent From a Bayesian perspective, one should compute the predictions from full
distribution of the hyperparameters, to take into account their correlations and
uncertainties. In a very recent work by the present authors \cite{mnras23},
$10^{4}$ random realizations were obtained from the full distribution of
hyperparameters. This exercise was undertaken for four kernels, to investigate if
the different covariance functions lead to significant changes in the final results.
However, no appreciable differences was found regardless of the kernel choice. Another
effort was carried out in \cite{adria2023} with similar conclusion.

\noindent It is expedient to abandon the assumption of a Dirac delta distribution
for the hyperparameters and propagate their non-zero uncertainties to the reconstructed
function. But, this entire procedure becomes computationally expensive. Besides, in
the case of constant prior mean, the likelihood is almost sharply peaked, so the
best-fit result becomes a good approximation. So, we compute the best-fit values of
the hyperparameters to determine the predictions from our GP model.

\noindent To reconstruct the comoving distance $d_C(z)$, its derivative and the corresponding
uncertainties, we consider the 1701 apparent magnitudes from Pantheon+ compilation of
SNIa \cite{scolnic2021}. We have removed those SN data that are contained in the host galaxies
of  SH0ES \cite{riess2021, brout2022}, in order to obtain results independent of them. The
observed 1624 apparent magnitudes $\tilde{m}$ for each SN Ia lightcurve as measured
on Earth depends on heliocentric $z_{\text{hel}}$ and Hubble diagram $z_{\text{hd}}$ redshifts.
We can rewrite the expression the apparent magnitudes in absence of peculiar motions, only in
terms of the redshift $z_{\text{hd}}$ as,
\begin{equation}
	m(z_{\text{hd}}) = \tilde{m}(z_{\text{hel}}, \, z_{\text{hd}}) - 5\, \log_{10}
	\left( \frac{1+z_{\text{hel}}}{1+z_{\text{hd}}}\right) = M + 25 + 5 \, \log_{10}
	\left[ \frac{d_L(z_{\text{hd}})}{1 \, \text{Mpc}} \right],
\end{equation}
where $M$ is peak absolute magnitude of the supernovae.

\noindent Following a similar prescription by Dinda \cite{dinda2022}, we generate the
function values $m(z)$, $m'(z)$ and their corresponding uncertainties at the CC redshifts
employing another GP. To scale down the drastic variance in the density of data with
redshift the reconstruction is carried out in $\ln(z)$. With these reconstructed $m(z)$
and $m'(z)$, we obtain the model-independent constraints on $M$ by minimizing the $\chi^2$
function,
\begin{equation}
	\chi^2 = {(H^{\text{CC}} - H^{\text{SN}})}^{\text{T}} \cdot {\mathlarger{\mathlarger
			{\mathlarger{\Sigma}}}}^{-1} (H^{\text{CC}} - H^{\text{SN}}) \, ,
\end{equation}
considering a uniform prior for $M \, \in \, \left[-21, -18\right]$ and $\Omega_{k0} h^2
\, \in \, [-0.5, 1]$ with $h = \frac{H_0}{100 \, \text{km Mpc$^{-1}$ s$^{-1}$}}$
	as the reduced, dimensionless, Hubble constant. Here, $H^{\text{CC}}$ are the CC $H(z)$ measurements,
$H^{\text{SN}}$ is the Hubble parameter derived from the SN data, and
$\mathlarger{\mathlarger{\Sigma}} \equiv \Sigma^{\text{CC}} + \Sigma^{\text{SN}}$
is the total covariance matrix. Therefore $H^{\text{SN}}$ is given by,
\begin{equation}
	H^{\text{SN}} = \left\{ \frac{(1+z)^2 \left[  c^2 \, (1+z)^2 + \Omega_{k0} \, H_0^2 \,
		d_L^2 \right]}{\left[ (1+z)d_L' - d_L \right]^2} \right\}^{\frac{1}{2}} ,
\end{equation}
where $d_L$ and $d_L'$ are the luminosity distances and its derivatives that can be computed
from the reconstructed $m$ and $m'$ values as,
\begin{eqnarray}
	d_L(z) &= 10^{\frac{1}{5} \left(m - M - 25\right)},  \label{eq:dLz} \\
	d_L'(z) &= m' \, \frac{\log{(10)}}{5} \,  10^{\frac{m- M -25}{5}}.
\end{eqnarray}
Plots for the marginalized constraints on $M$, obtained via MCMC analysis, is shown
in Fig. \ref{fig:M_contour}. Note that, although these constraints are model-independent
they implicitly assume the validity of the cosmic distance duality relation.

\noindent The best fit values and 1$\sigma$ uncertainties for the parameters $M$,
$\Omega_{k0} \, h^2$ for the CC calibrated SN data, and $r_d$ from the CC calibrated
BAO data is given in Table \ref{tab:constraints}.

\noindent With the respective constraints on $M$, we can generate the comoving distance
data using Eq. (\ref{eq:dz}) in Eq. (\ref{eq:dLz}). Finally, we can carry out a GP
reconstruction of the functions $d_C(z)$, $d_C'(z)$ and the respective uncertainties. A
triangle plot showing the marginalized posteriors of the GP hyperparameters for $d_C(z)$
is shown in the right panel of Fig. \ref{fig:hyper_plot}.

\noindent To determine the most suitable kernel we select the associated GP model
that gives the minimum $\chi^2$ between the observations vs reconstructed values at
the training redshifts. Table \ref{tab:chi2r} gives the reduced best-fit $\chi^2$
obtained with different data sets for the six kernels. We also compare the effects
resulting from the remaining best-fit kernel choices. We find that the Mat\'{e}rn
$\nu=5/2$ (M52) kernel performs better in comparison to the others.
As the M52 kernel has the largest posterior distribution in all situations, it indicates
that the GP reconstructed functions employing the M52 kernel seems to be the closest
to the real data and hence can be considered as a better model. Recently, Zhang {\it
et al} \cite{zhang2023} arrived at similar results using the ABC Rejection algorithm
for kernel selection.

\noindent We show the plots for the reconstructed functions, $H(z)$, $H'(z)$, from
the CC and combined CC+BAO data sets in Figs. \ref{fig:H_rec} and \ref{fig:H2_rec}
respectively. The reconstructed profile of $d_C(z)$ and $d_C'(z)$ from the Pantheon+
SN data is shown in Fig. \ref{fig:dc_rec}. We also plot the correlations between the
reconstructed functions and their derivatives simultaneously. All the figures above
have been generated employing the Mat\'ern 5/2 covariance function.

\begin{figure*}[t!]
	\begin{minipage}{\textwidth}
		\begin{minipage}{0.325\textwidth}
			\includegraphics[width=\textwidth]{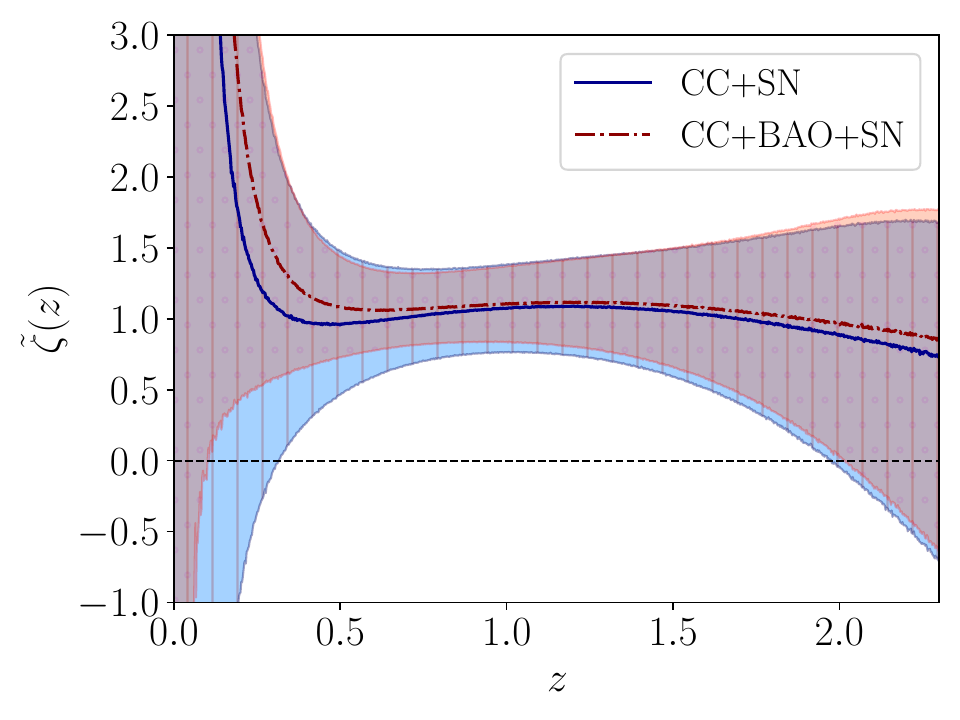}
			\caption*{(a) RBF}
		\end{minipage}%
		\begin{minipage}{0.325\textwidth}
			\includegraphics[width=\textwidth]{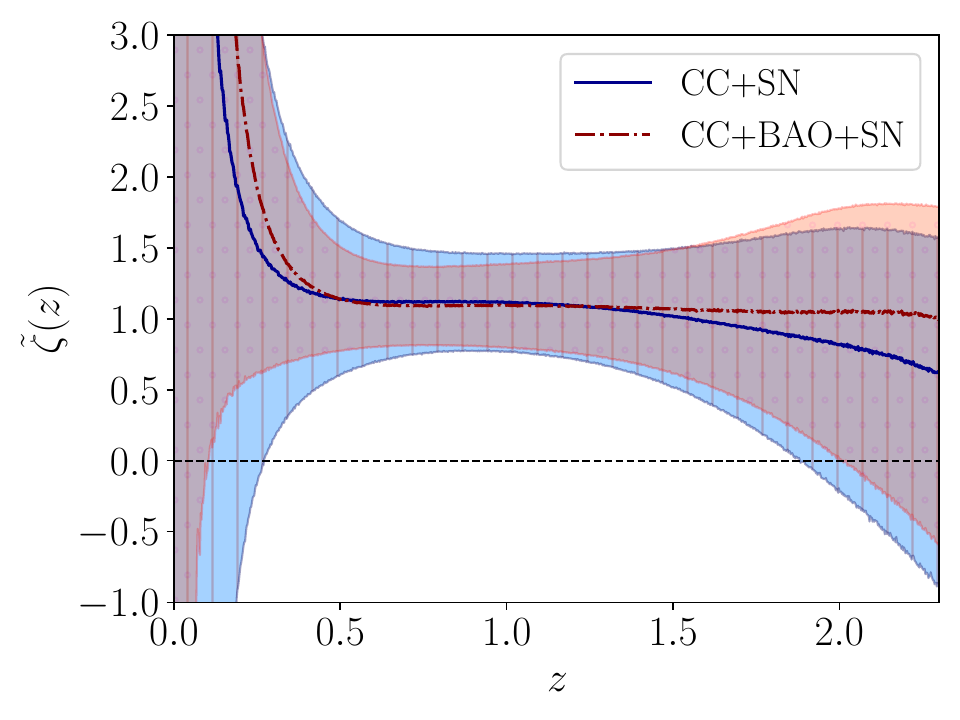}
			\caption*{(b) CHY}
		\end{minipage}%
		\begin{minipage}{0.325\textwidth}
			\includegraphics[width=\textwidth]{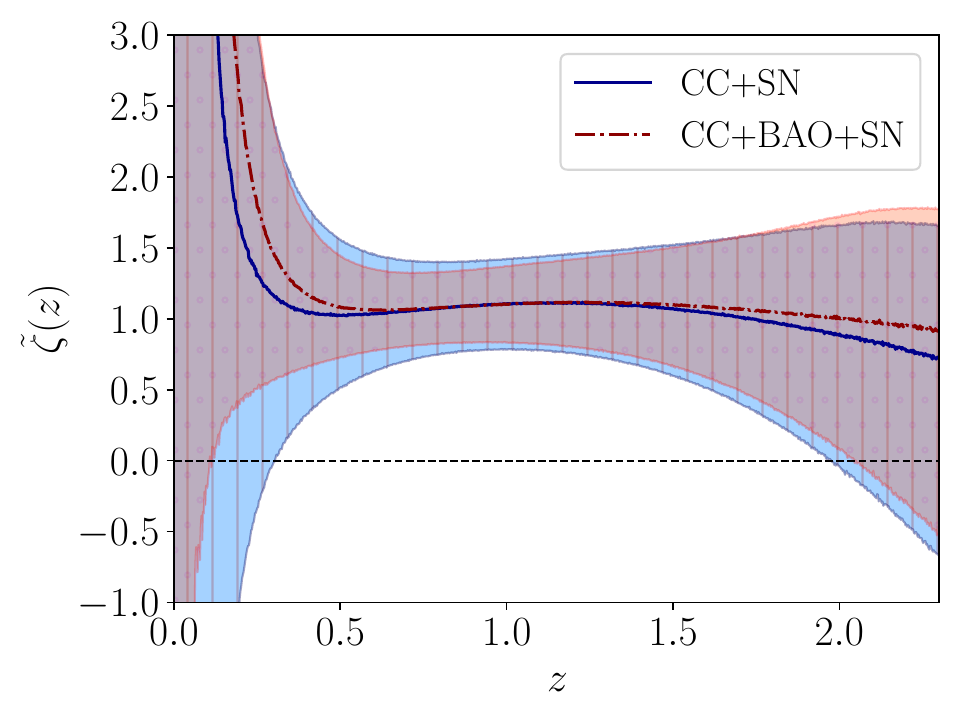}
			\caption*{(c) RQD}
		\end{minipage}%
	\end{minipage}%
	\vskip 0.2cm
	\begin{minipage}{\linewidth}
		\begin{minipage}{0.325\textwidth}
			\includegraphics[width=\textwidth]{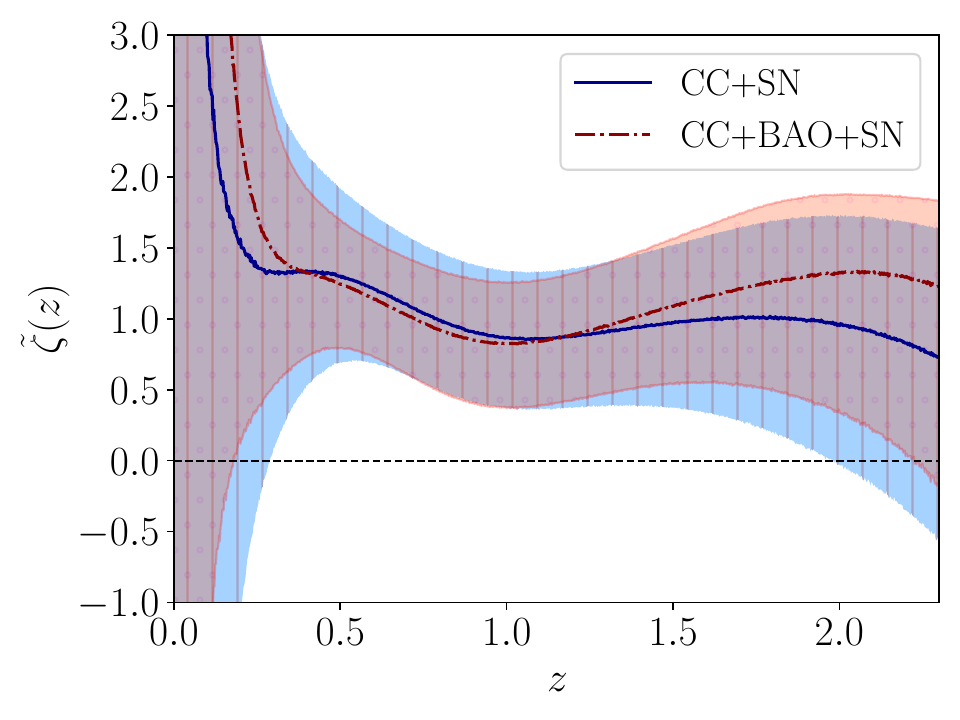}
			\caption*{(d) M52}
		\end{minipage}%
		\begin{minipage}{0.325\textwidth}
			\includegraphics[width=\textwidth]{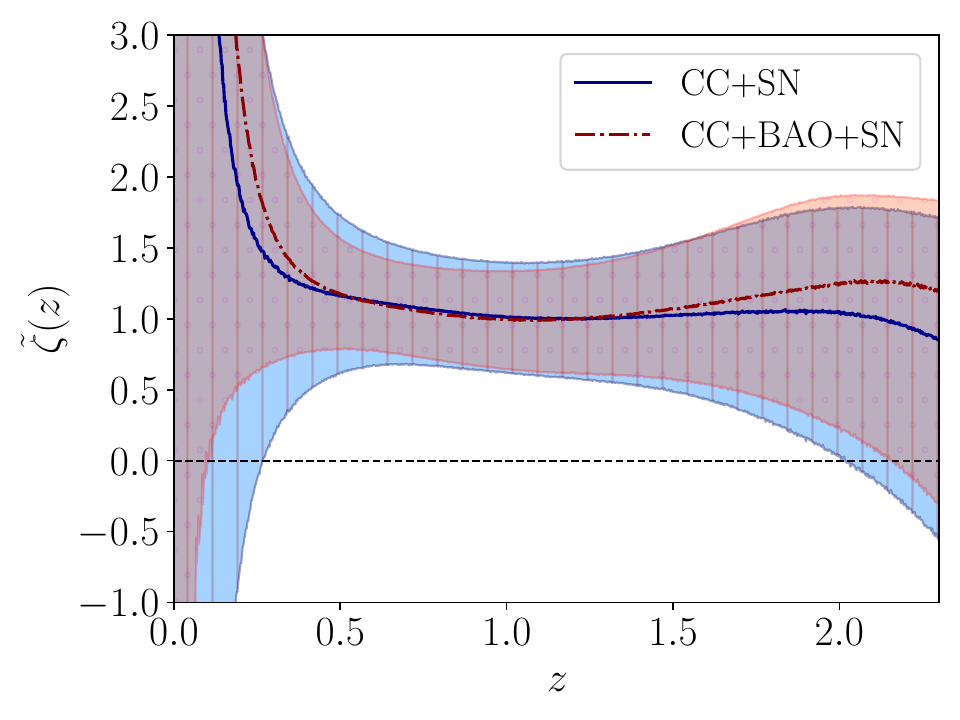}
			\caption*{(e) M72}
		\end{minipage}%
		\begin{minipage}{0.325\textwidth}
			\includegraphics[width=\textwidth]{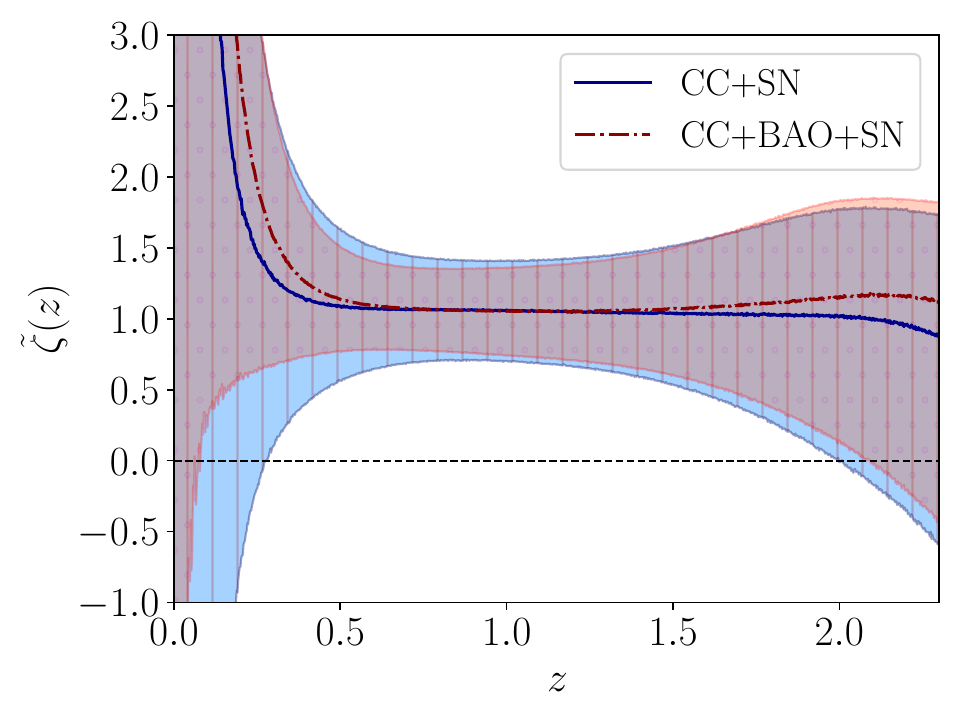}
			\caption*{(f) M92}
		\end{minipage}%
	\end{minipage}%
	\hfill 	
	\caption{Reconstructed $\tilde\zeta$, after using Eq. \eqref{eq:zeta_recon}
		from CC+SN (solid lines) and CC+BAO+SN (dash-dot lines) data set combinations for
		different kernel choices. The shaded regions with `|' and `$\cdot$' hatches represent
		the 1$\sigma$ confidence levels for the CC+SN and CC+BAO+SN combinations respectively.
		$\tilde\zeta$ diverges when $z$ approaches zero because so does $d_{C}$.}
	\label{fig:zeta_recon}
\end{figure*}

\section{Results \label{sec:5}}

\noindent Finally, with the reconstructed functions $H(z)$, $H'(z)$, $d_C(z)$ and $d_C'(z)$,
we now derive the evolution of $\zeta$, defined by Eq. \eqref{eq:zeta_recon} as a function of
the cosmological redshift. Alternatively, we can substitute the constraints obtained on  $\Omega_{k0} h^2$,
given in Table \ref{tab:constraints}, to arrive at $\Omega_{k0} H_0^2$ and
directly use Eq. \eqref{eq:zeta} to reconstruct $\zeta(z)$.

\noindent However, in equations \eqref{eq:zeta} and \eqref{eq:zeta_recon}, $\zeta$ has units of the Hubble factor.
This is made clear by writing Eq. \eqref{eq:zeta} and \eqref{eq:zeta_recon} as,
\begin{equation}
	\zeta(z) \equiv H_{0} \left[E'(z) -  \Omega_{k0} \frac{1+z}{E(z)} \right] \geq 0
	\label{eq:equivalent1}
\end{equation}
or, equivalently,
\begin{equation}
	\zeta(z) \equiv  H_{0} \left[ E'(z) - \frac{E^{2}(z) \, D'^{2}(z) - 1}{D^{2}(z)} \, \frac{1+z}{E(z)} \right] \geq 0.
	\label{eq:equivalent2}
\end{equation}

\noindent Obviously, it is expedient to have a dimensionless representation. So, we use
	$\tilde{\zeta} \equiv \zeta/H_0$ with $H_0 = 73.3 \text{ km Mpc$^{-1}$  s$^{-1}$}$, the central SH0ES value for
the Hubble constant \cite{riess2021}. It deserves
mention that this assumption  serves the purpose of a rescaling just to make the reconstructed
quantity dimensionless. So, our final results are independent of this choice.

\noindent Figure \ref{fig:zeta_recon} shows plots of the evolution of $\tilde{\zeta}$
along with their 1$\sigma$ confidence levels for the CC+SN and CC+BAO+SN data set combinations. Here, we show the results obtained
with different reconstruction kernels for comparison and better illustration. Note that the reconstructed $\tilde{\zeta}$
from Eq. \eqref{eq:zeta_recon} has a divergence since $d_C(z = 0) = 0$ brings a singularity at $z = 0$.

\begin{figure*}[t!]
	\begin{minipage}{\textwidth}
		\begin{minipage}{0.325\textwidth}
			\includegraphics[width=\textwidth]{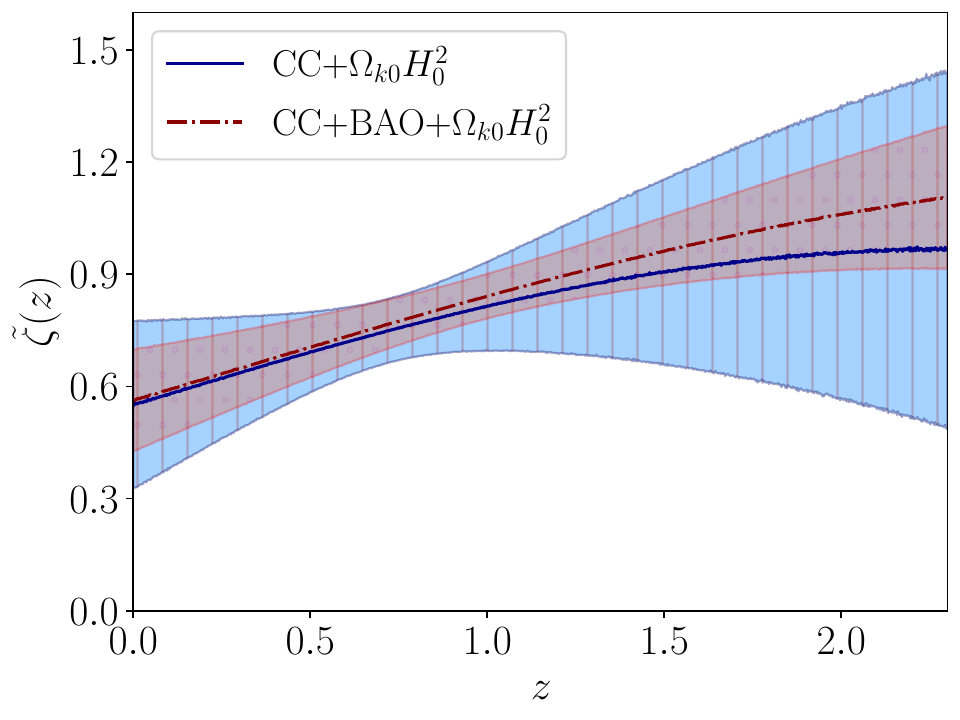}
			\caption*{(a) RBF}
		\end{minipage}%
		\begin{minipage}{0.325\textwidth}
			\includegraphics[width=\textwidth]{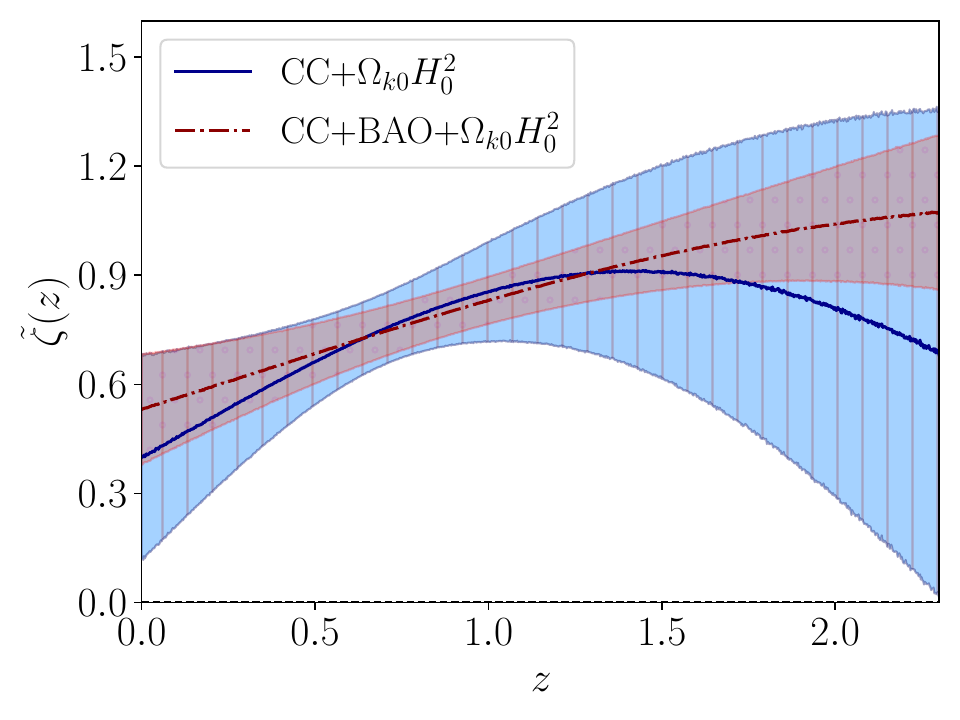}
			\caption*{(b) CHY}
		\end{minipage}%
		\begin{minipage}{0.325\textwidth}
			\includegraphics[width=\textwidth]{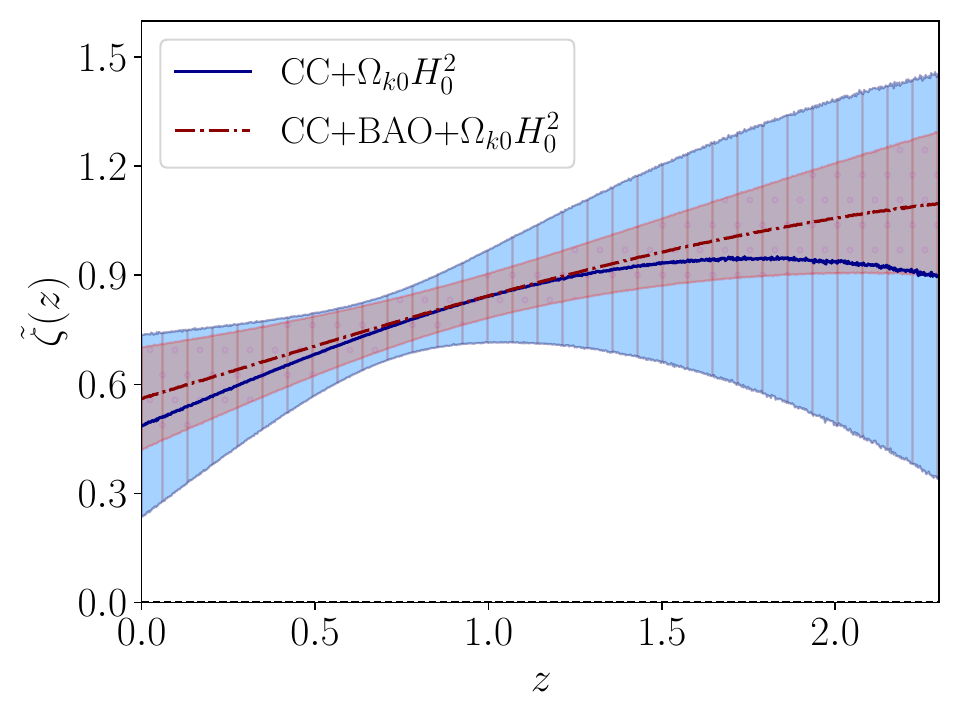}
			\caption*{(c) RQD}
		\end{minipage}%
	\end{minipage}%
	\vskip 0.2cm
	\begin{minipage}{\linewidth}
		\begin{minipage}{0.325\textwidth}
			\includegraphics[width=\textwidth]{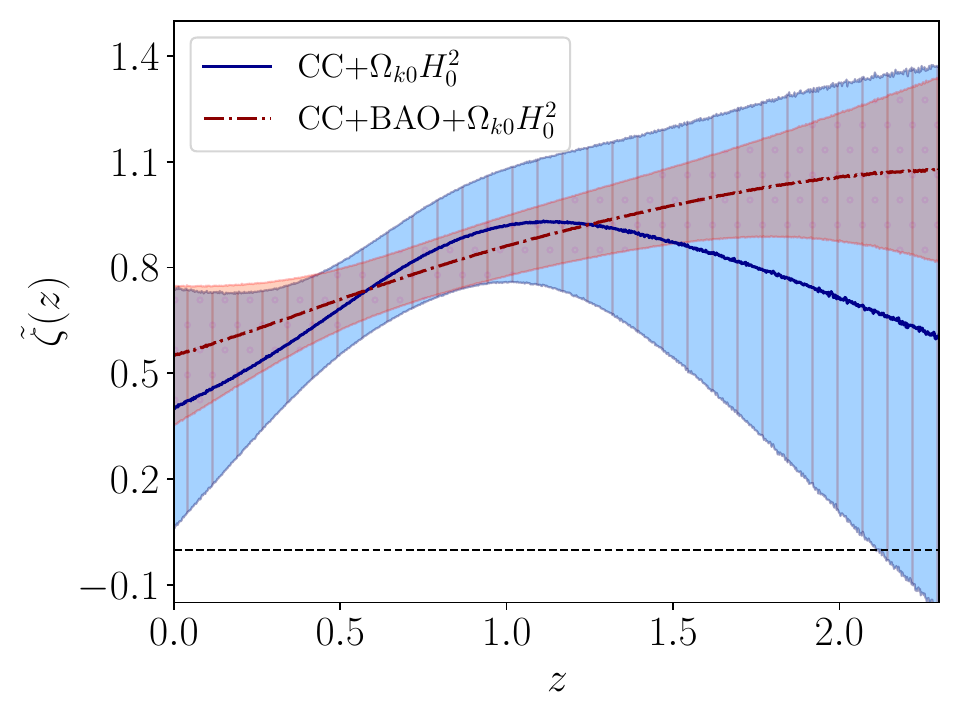}
			\caption*{(d) M52}
		\end{minipage}%
		\begin{minipage}{0.325\textwidth}
			\includegraphics[width=\textwidth]{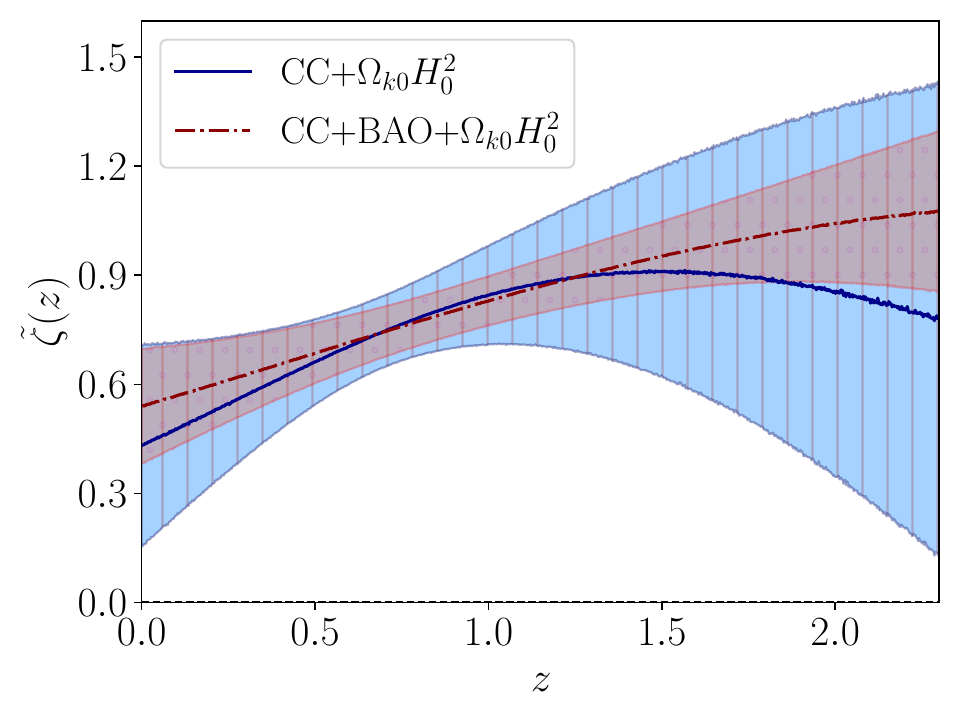}
			\caption*{(e) M72}
		\end{minipage}%
		\begin{minipage}{0.325\textwidth}
			\includegraphics[width=\textwidth]{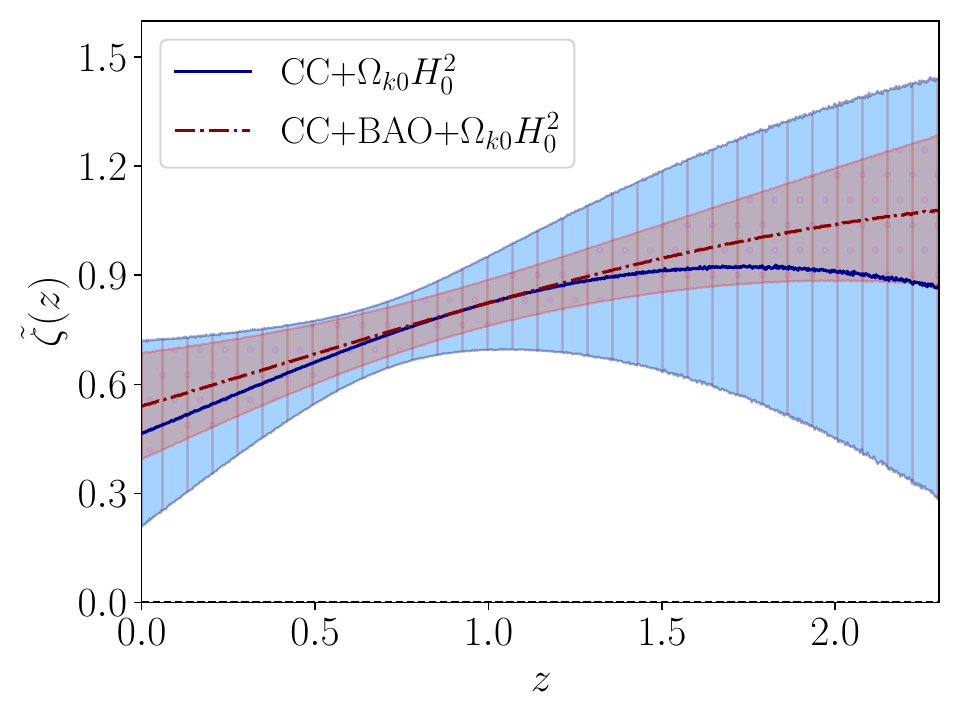}
			\caption*{(f) M92}
		\end{minipage}%
	\end{minipage}%
	\hfill 	
	\caption{Reconstructed $\tilde{\zeta}$, derived using Eq. \eqref{eq:zeta} from
		CC+$\Omega_{k0}H_0^2$ (solid lines) and CC+BAO+$\Omega_{k0}H_0^2$ (dash-dot lines)
		combinations for different kernel choices. The shaded regions with `|' and `$\cdot$'
		hatches represent the 1$\sigma$ confidence levels for the CC+$\Omega_{k0}H_0^2$ and
		CC+BAO+$\Omega_{k0}H_0^2$ combinations respectively.}
	\label{fig:zeta}
\end{figure*}

\noindent We also plot for the redshift evolution of $\tilde{\zeta}$ using Eq. \eqref{eq:zeta} along with its
1$\sigma$ confidence levels in Fig. \ref{fig:zeta} for the CC+$\Omega_{k0}H_0^2$ and CC+BAO+$\Omega_{k0}H_0^2$
combinations. We take into account the correlations between the reconstructed functions and their derivatives,
and compute the uncertainties associated with $\tilde{\zeta}$ via the Monte Carlo error propagation rule.

\noindent As is apparent in these figures neither the solid nor the dashed-dot lines  ever cross
	the horizontal axis $\tilde{\zeta} =0$.

\noindent Before closing this section, it is worth mentioning  that our findings severely 
constrain universe models in which phantom dark energy dominates the expansion. In such models 
the expansion is super-accelerated ($\dot{H} >0$), i.e.  $q < -1$. While observational data do 
not exclude them \cite{aghanim2020}, and new proposals have been made recently \cite{cruz2022}, 
equation \eqref{1plusq} and the fact that the data strongly suggest that $\Omega_{k0}$ is 
either zero or very close to it \cite{aghanim2020}, it follows that  it is extremely likely 
that $ q \geq -1$  and that  $\tilde{\zeta} >0$. Although the results summarized in  Fig. 
\ref{fig:zeta_recon} do not rule out  entirely the possibility $\tilde{\zeta} < 0 $ in the 
redshift range $0 < z < 2$, the results summarized in Fig. \ref{fig:zeta} do. In addition, 
phantom models face serious difficulties from the theoretical side since, their energy density  
being unbounded from below, they do not appear stable either from quantum consideration 
\cite{cline2004} or even classically \cite{dabrowski2015}.

\section{Concluding Remarks \label{sec:6}}

\noindent Our study, based solely on Einstein gravity and the assumption that the Universe is
homogeneous and isotropic, backs previous analysis about the question, ``is the second law of
thermodynamics satisfied at cosmic scales?''
After making use of the history of thee Hubble function in the redshift range $0 < z \leq 2$  we
find the answer in the affirmative. Indeed, as Figs. \ref{fig:zeta_recon} and  \ref{fig:zeta} show, $\tilde{\zeta}$
stays positive in the said range; i.e., the inequality $1+q \geq \Omega_{k}$ is fulfilled in that interval
irrespective of whether $\Omega_{k}$ is positive, negative or nil. This means that the history of the Hubble
function  tells us that the area of the apparent horizon does not decrease with expansion. In other words,
the mentioned history is compatible with second law of thermodynamics at cosmic scales as it was 
suggested by quite different methods in \cite{mnrs2019} and \cite{prd2018}. This result is most reassuring; it would be 
weird that this law, being valid at terrestrial and astrophysical scales, would fail at cosmic scales. 
Finally, as a by product, this work strengthens our belief that the Universe was never dominated by phantom
	fields in the considered redshift range.

\acknowledgments

PM thanks ISI Kolkata for financial support through Research Associateship.

\label{lastpage}

\end{document}